\documentclass[aps,pre,reprint,superscriptaddress]{revtex4-1}

\usepackage{graphicx}
\usepackage{amssymb}
\usepackage{amsmath}
\usepackage{bm}
\usepackage{color}
\usepackage{hyperref} 

\graphicspath{{./figures/}}

\begin{document}
\title{From local to hydrodynamic friction in Brownian motion:\\ A multiparticle collision dynamics simulation study}

\author{Mario Theers}
\email{m.theers@fz-juelich.de}
\affiliation{Theoretical Soft Matter and Biophysics, Institute for
Advanced Simulation and Institute of Complex Systems,
Forschungszentrum J\"ulich, D-52425 J\"ulich, Germany}
\author{Elmar Westphal}
\email{e.westphal@fz-juelich.de}
\affiliation{Peter Gr\"unberg Institute and J\"ulich Centre for Neutron
Science, Forschungszentrum J\"ulich, D-52425 J\"ulich, Germany}
\author{Gerhard Gompper}
\email{g.gompper@fz-juelich.de}
\affiliation{Theoretical Soft Matter and Biophysics, Institute for
Advanced Simulation and Institute of Complex Systems,
Forschungszentrum J\"ulich, D-52425 J\"ulich, Germany}
\author{Roland G. Winkler}
\email{r.winkler@fz-juelich.de}
\affiliation{Theoretical Soft Matter and Biophysics, Institute for
Advanced Simulation and Institute of Complex Systems,
Forschungszentrum J\"ulich, D-52425 J\"ulich, Germany}
\date{\today}

\begin{abstract}
The friction and diffusion coefficients of rigid spherical colloidal particles dissolved in a fluid are determined from velocity and force autocorrelation functions by mesoscale hydrodynamic simulations.
Colloids with both slip and no-slip boundary conditions are considered, which are embedded in fluids modelled by multiparticle collision dynamics (MPC)  with and without angular momentum conservation.
For no-slip boundary conditions, hydrodynamics yields the well-known Stokes law,
while for slip boundary conditions the lack of angular momentum conservation leads to a reduction of the hydrodynamic friction coefficient compared to the classical result. The colloid diffusion coefficient is determined by integration of the velocity autocorrelation function, where the numerical result at
shorter times is combined with the theoretical hydrodynamic expression for longer times.
The suitability of this approach is confirmed by simulations of sedimenting colloids.
In general, we find only  minor deviations from the Stokes-Einstein relation, which even disappear for larger colloids. Importantly,
for colloids with slip boundary conditions, our simulation results contradict the frequently assumed additivity of local and hydrodynamic diffusion coefficients.
\end{abstract}
\maketitle

\section{Introduction}

The dynamics of colloids in solution is governed by hydrodynamic interactions, a fact which is very well established \cite{kim:13,dhon:96,happ:12}. Yet, a full account of the long-range nature of the hydrodynamic interactions in computer simulations of large systems of colloidal dispersions is still challenging. Recently developed mesoscale hydrodynamic simulation approaches, such as the multiparticle collision dynamics (MPC) method \cite{male:99,kapr:08,gomp:09}, are very valuable to study colloidal dispersions, with the strong length- and time-scales separation between the colloid and fluid degrees of freedom.
The particle-based MPC approach has been shown to correctly reproduce the hydrodynamic properties of embedded colloids or polymers \cite{gomp:09,huan:13}.
Various approaches have been proposed for the embedding of a spherical colloid into the MPC fluid \cite{lee:04,padd:05,whit:10,pobl:14}.
The corresponding colloid-fluid coupling governs the colloid's velocity or force autocorrelation function (VACF, FACF). Here, two different time regimes are typically identified: a short-time regime of uncorrelated fluid-particle motion and local interactions, characterized by molecular chaos, and a long-time regime with strong  hydrodynamic correlations. The first regime is typically dominated by ballistic fluid-colloid collisions and is described by the Enskog gas theory \cite{odel:75,bern:76.1,whit:10}; in a more general sense, we will denote this regime as {\em local} regime. As a consequence, the VACF decays exponentially for short times with a characteristic time given by the ratio of the local friction coefficient and the
colloid mass \cite{padd:05,goet:10,whit:10,belu:11,pobl:14}. On longer time scales, the VACF follows the prediction of hydrodynamic theory (Navier-Stokes) and displays an algebraic long-time tail \cite{alde:70,padd:05,goet:10,whit:10,belu:11,huan:12,pobl:14}.

The separation between local and hydrodynamic time scales affects the frictional and consequently the diffusive behavior of a colloid. Indeed,
several studies suggest that the total colloid diffusivity is a sum of a hydrodynamic and local diffusion coefficient \cite{lee:04,padd:06,whit:10}.
However, a detailed and decisive study of the relevance of the various contributions for a colloid embedded in a MPC fluid is still missing.

The standard, most often applied implementation of MPC \cite{kapr:08,gomp:09} does not conserve angular momentum, which can give rise to unphysical torques \cite{goet:07}, and, in case of (partial) slip boundaries, yields flow fields, which deviate from those theoretically predicted \cite{yang:15}.
In Ref.~\cite{yang:15}, it has been found in particular that the lack of angular momentum conservation combined with (partial) slip  boundary conditions leads to a reduction of the friction coefficient compared to that of a fluid with angular momentum conservation. This result is surprising, especially since it was not reported before in studies where the friction of colloids with slip boundary conditions was determined \cite{lee:04,padd:06,whit:10}.
The reason may lie in the rather involved interpretation of simulated friction coefficients.
As mentioned above, the friction measured in MPC simulations is not only determined by hydrodynamics, but also by short-time local processes.
Additionally, periodic boundary conditions, typically employed in simulations, affect the frictional behavior. The importance of the various contributions has been addressed before \cite{lee:04,padd:06,whit:10}, but the validity of Stokes' law for systems with slip boundary conditions has been presumed. In the light of the modified Stokes law \cite{yang:15}, the frictional behavior needs to be reconsidered to correctly identify the hydrodynamic and local contributions.

In this paper, we generalize the steady-state considerations for the friction coefficient of Ref.~\cite{yang:15} and determine its full frequency dependence.
We verify the reduction of the hydrodynamic friction of non-angular-momentum-conserving MPC methods compared to angular momentum conserving ones by measuring force as well as velocity autocorrelation functions. Moreover, the relative contributions of local and hydrodynamic friction are determined by the force autocorrelation function.
We determine the diffusion and friction coefficients for colloids of different radii by integration of the autocorrelation function (Green-Kubo) and complement the analysis by sedimentation studies.
This enables us to examine the additivity of various contributions to the diffusion coefficient.
We find that the hydrodynamic frictional contribution dominates by far for slip as well as no-slip boundary conditions.
The local friction only yields a small additional contribution for small colloid radii.
For colloids with no-slip boundaries, the sum of the hydrodynamic and local diffusion coefficient yields a good approximation for the total diffusion coefficient. However, for slip boundary conditions this additivity does not apply.

The paper is structured as follows. In section \ref{sec:MPC} we present the multiparticle collision dynamics method and the colloid-solvent coupling.
The relevant fluctuation-dissipation theorems for Brownian motion as well as hydrodynamic and local friction are briefly discussed in section \ref{sec:Brownian}.
Simulation results are presented in section \ref{sec:Simulations}; section \ref{sec:Conclusions} summarizes our findings.
Finally, we include appendixes, in which analytical formula
are derived for hydrodynamic and Enskog friction, the latter being a special case of local friction
(see appendix \ref{sec:FrequencyDependentHydrodynamicFriction} and \ref{sec:EnskogFriction}, respectively).

\section{Multiparticle collision dynamics} \label{sec:MPC}

\subsection{Algorithm}

The MPC solvent is modeled by $N$ point particles with mass $m$, positions
${\bm r}_i$, and velocities ${\bm v}_i$ ($i=1,\cdots,N$), contained in a cubic simulation box with periodic boundary conditions.
The discrete time dynamics consists of a streaming step, for the collision-time interval $h$, and a subsequent instantaneous collision.
In the ballistic streaming step, the particle positions are
updated via
\begin{align} \label{EqSolventStreaming}
 {\bm r}_i(t+h)={\bm r}_i(t)+ h {\bm v}_i(t).
\end{align}
In the collision step, the simulation
box is partitioned into cubic collision cells of length $a$, in which stochastic
multiparticle collisions are performed. For these collisions, different approaches
have been developed  \cite{male:99,alla:02,gomp:09,nogu:07,thee:15}.
In the stochastic-rotation-dynamics version (MPC-SRD) \cite{male:99,kapr:08,gomp:09}, the relative
velocity of each particle, with respect to the center-of-mass velocity of the
cell, is rotated by a fixed angle $\alpha$ around a randomly oriented axis, independent for each cell, which yields the velocities
\begin{align} \label{MPC_rotation}
 {\bm v}_i (t+h)={\bm v}_{cm}(t)+{\bm R}(\alpha)({\bm v}_i(t)- {\bm v}_{cm}(t)).
\end{align}
Here, ${\bm v}_i(t)$ and ${\bm v}_i(t+h)$ are the velocities before and after the collision, respectively.
${\bm R}(\alpha)$ is the rotation matrix,
\begin{align}
{\bm v}_{cm}=\frac{1}{N_c} \sum_{j \in cell} {\bm v}_j
\end{align}
is the center-of-mass velocity, and $N_c$ the total number of particles in the cell of particle $i$.
To maintain Galilean invariance, a random shift of the collision grid is performed in every collision step \cite{ihle:01,ihle:03}.
The update in Eq. (\ref{MPC_rotation}) conserves energy and momentum. A canonical ensemble is achieved by a suitable thermostat \cite{hech:05,huan:10.1}.
Throughout this paper, we apply the local Maxwell-Boltzmann-scaling (MBS) thermostat,
where the relative particle velocities in each cell are scaled by a factor determined from the Gamma distribution of cell kinetic energies \cite{huan:10.1,huan:15}.

On length scales larger than a collision cell, the MPC fluid obeys the Navier-Stokes equations \cite{male:99,ihle:09,huan:12}.
However, in contrast to real fluids, the stress tensor of the above MPC-SRD implementation is non-symmetric, since angular momentum is not conserved during the collision step \cite{ihle:05,pool:05,yang:15}.
A symmetric stress tensor follows by an extension of Eq.
(\ref{MPC_rotation}), which ensures angular momentum conservation on a cell level \cite{nogu:08}.
The velocity update then reads as \cite{nogu:08,thee:15}
\begin{align} \label{Eq:AMC_vel_update}
\begin{split}
 {\bm v}_i(t+h) &=  ~{\bm v}_{cm}+{\bm R}(\alpha){\bm v}_{i,c} \\
 &- {\bm r}_{i,c} \times \Big[ m {\bm I}^{-1} \sum_{j \in cell}\left\{{\bm r}_{j,c}\times \left({\bm v}_{j,c}-{\bm R}(\alpha) {\bm v}_{j,c} \right) \right\} \Big],
\end{split}
\end{align}
where ${\bm r}_{i,c}={\bm r}_{i}-{\bm r}_{cm}$ is the particle position relative to their center-of-mass ${\bm r}_{cm}$ of a cell,
${\bm I}$ is the moment-of-inertia tensor of the particles in the center-of-mass reference frame, and
${\bm v}_{i,c}={\bm v}_{i}-{\bm v}_{cm}$.
It is not necessary that the update in Eq. (\ref{Eq:AMC_vel_update}) conserves energy, since we apply the MBS-thermostat subsequently. MPC-SRD with and without angular momentum conservation are referred to as MPC-SRD+a and MPC-SRD-a, respectively.
The stress tensor of the MPC fluid is of the general form \cite{thee:15}
\begin{align} \label{StressTensorGen}
  {\bm \sigma} =-p ~{\bm E} +\eta_1 {\bm \nabla} {\bm v}^T + \eta_2 ({\bm \nabla} {\bm v}^T)^T + \eta_3 ({\bm \nabla} \cdot {\bm v}) ~ {\bm E},
\end{align}
where $p(\bm{r},t)$ is the pressure field, $\bm{v}(\bm{r},t)$ is the velocity field, $\eta_1$, $\eta_2$, and $\eta_3$ are the viscosity parameters, and $\bm E$ is the unit matrix.
The stress tensor of MPC-SRD-a is asymmetric with $\eta_1=\eta^k$, $\eta_2= \eta = \eta^c+\eta^k$, and $\eta_3=-2\eta_1/3$ \cite{pool:05,thee:15}, where $\eta$ is the shear viscosity with its kinetic  $\eta^k$ and collisional $\eta^c$ contribution.
In case of MPC-SRD+a, the stress tensor is symmetric and $\eta_1=\eta_2=\eta$, $\eta_3=-2\eta/3+\eta^V$, where  $\eta^V=(\eta_1+\eta_2+3\eta_3)/3$ is the bulk viscosity \cite{thee:15}.
Note that the viscosities $\eta^k$ and $\eta^c$ depend on the presence or absence of angular momentum conservation \cite{nogu:08}, but $\eta^V$ is equal for MPC-SRD-a and MPC-SRD+a, since the bulk viscosity of MPC-SRD+a is determined by the collisional viscosity of MPC-SRD-a \cite{thee:15}.

\subsection{Colloid-solvent coupling} \label{SubSec:Collod-Solvent-Coupling}

A spherical colloid of radius $R$, mass $M$, velocity $\bm{u}$, and angular velocity $\bm{\Omega}$ can be coupled to the solvent by elastic collisions during the streaming step \cite{whit:10,padd:05}.
Each of these collisions transfers a linear momentum ${\bm J}_i$, while conserving the total linear
and angular momentum as well as energy. Denoting post-collisional quantities by a prime, we may write \cite{whit:10,yang:14}
\begin{align}
 {\bm v}_i'&={\bm v}_i-{\bm J}_i/m , \\
 {\bm u}'&={\bm u}+{\bm J}_i/M , \\ \label{eq:angular}
 {\bm \Omega}'&={\bm \Omega}+ R ({\bm n}_i \times {\bm J}_i)/I ,
\end{align}
which ensures conservation of linear and angular momentum for every choice of ${\bm J}_i$.
Here, ${\bm v}_i$ is the velocity of the colliding MPC particle, $I=\chi M R^2=(2/5)M R^2$ is the colloid's moment of inertia, and ${\bm n}_i=(\bm{r}_i-\bm{C})/|\bm{r}_i-\bm{C}|$ is the unit vector between the particle position $\bm{r}_i$  and the colloid center $\bm{C}$.
We define the relative velocity
\begin{align} \label{Eq:def_v_bar}
  \bar{\bm v}_i={\bm v}_i-{\bm u}-{\bm \Omega} \times R {\bm n}_i
\end{align}
between the MPC particle and the closest point on the colloid surface.
By using energy conservation, we obtain two solutions for $\bm{J}_i$, which we distinguish by the parameter $\Gamma \in \{0,1\}$ \cite{whit:10,yang:14}, hence,
\begin{align} \label{MomentumTransferCollisionsColloid}
 {\bm J}_i=2 \mu \bar{{\bm v}}_{i,n}+2(1-\Gamma)\mu \frac{M \chi}{\mu+M \chi}\bar{{\bm v}}_{i,t}.
\end{align}
Here, $\mu=m M/(m+M)$ is the reduced mass and the indices $n$ and $t$ indicate the normal and tangential components,
i.e., $\bar{\bm{v}}_{i,n}=\bm{n}_i\bm{n}_i^T \bar{\bm{v}}_i$ and $\bar{\bm{v}}_{i,t}=(1-\bm{n}_i\bm{n}_i^T) \bar{\bm{v}}_i$.
For $M \gg m$, the choice $\Gamma=1$ corresponds to a specular reflection, and hence slip boundary conditions, while $\Gamma=0$ corresponds to
bouncing-back and thus no-slip boundary conditions. As a generalization, by the choice $\Gamma \in (0,1)$  partial slip can be modeled \cite{whit:10,hu:15}.

The MPC fluid-colloid collisions are performed in a coarse-grained manner. At first, the solvent particles as well as the colloid move ballistically
according to Eq. (\ref{EqSolventStreaming}) and
\begin{align} \label{EqColloidStreaming}
  {\bm C}(t+h)={\bm C}(t)+h {\bm u}(t) ,
\end{align}
respectively.
At second, each solvent particle $i$ with $|{\bm r}_i(t+h)-{\bm C}(t+h)|^2< R^2$ is moved back in time by $(h-h_i)$,
where $h_i$ is defined by $|{\bm r}_i(t)-{\bm C}(t) +h_i({\bm v}_i-{\bm u})|^2=R^2 $.
Each of those particles will then collide with a virtual colloid at position ${\bm C}(t) +h_i{\bm u}(t)$, with velocity ${\bm u}(t)$ and angular velocity ${\bm \Omega}(t)$,
transfer momentum ${\bm J}_i$ and subsequently move with its corrected velocity $\bm v_i'$ for the time $(h-h_i)$.
Hence, the streaming step Eq.~(\ref{EqSolventStreaming}) of a MPC particle interacting with a colloid is changed and consists of two parts, streaming before---with velocity $\bm v_i$---and after---with velocity $\bm v_i'$---the collision. Additionally, in the MPC collision Eq.~(\ref{MPC_rotation}), the velocity $\bm v_i'$ has to be used.   At third, the colloid's translational and rotational velocities are updated via
\begin{align} \label{EqColloidMPC_v}
{\bm u}(t+h) & ={\bm u}(t)+\sum_i {\bm J}_i/M , \\ \label{EqColloidMPC_o}
{\bm \Omega} (t+h) & ={\bm \Omega}(t) + R \sum_i ({\bm n}_i \times {\bm J}_i)/I .
\end{align}

In the case of no-slip boundaries, additionally MPC ghost particles \cite{lamu:01} are randomly distributed inside the colloid at each time step, which introduces an additional colloid-solvent interaction in the MPC collision step, and reduces the amount of slip.
After each MPC collision step, the colloid velocities are updated via
\begin{align}
  \bm{u}&=\bm{u}(t+h)+\frac{1}{M} \sum_i \bm{J}^{(g)}_i,  \\
  \bm{\Omega} &=\bm{\Omega}(t+h)+\frac{1}{I} \sum_i (\bm{r}^{(g)}_i -\bm{C}) \times \bm{J}^{(g)}_i,
\end{align}
where $\bm{J}^g_i$ denotes the change of momentum of the ghost particle $i$ at position $\bm{r}^{(g)}_i$ and $\bm u (t+h)$  and $\bm{\Omega}(t+h)$ are the velocities of Eqs.~(\ref{EqColloidMPC_v}) and (\ref{EqColloidMPC_o}).
For the treatment of ghost particles in case of partial-slip boundary conditions, see Refs. \cite{whit:10,hu:15}

Since the MPC algorithm is highly parallel, we execute simulations on a Graphics Processing Unit (GPU) for a high performance gain \cite{west:14}.

\section{Brownian motion} \label{sec:Brownian}

\subsection{Fluctuation dissipation relations}

The equations of motion of a colloidal Brownian particle of mass $M$ and velocity ${\bm u}(t)$ in a viscous fluid, experiencing a retarded friction force together with a random force ${\bm K}(t)$, are given by
\begin{align} \label{Eq_motion_Brownian_general}
 M \frac{d {\bm u}}{dt}=-\int_0^t dt'~ \gamma(t-t') {\bm u}(t')+{\bm K}(t).
\end{align}
The friction kernel $\gamma(t)$ obeys $\gamma(t)=0 ~ \forall~  t<0$ due to causality, but is not specified beyond that at this stage.
By multiplying Eq.~(\ref{Eq_motion_Brownian_general}) by $\bm{u}(0)$, averaging over the random forces,  and assuming $\langle {\bm K}(t) \cdot {\bm u}(0) \rangle=0$, we find the governing equation
\begin{align} \label{Eq:IntegrodifferentialEquation_for_C_u}
  M\frac{d C_u(t)}{dt}=-\int_0^t dt'~ \gamma(t-t') C_u(t'),
\end{align}
for the velocity autocorrelation function
\begin{align}
  C_u(t)=\frac{1}{3} \langle \bm{u}(t) \cdot \bm{u}(0)\rangle.
\end{align}
To solve Eq.~(\ref{Eq:IntegrodifferentialEquation_for_C_u}), we perform a Laplace transformation \cite{evan:08}, with $\tilde{f}(z)=\int_0^\infty dt ~ e^{-z t} f(t)$.
The resulting algebraic equation yields
\begin{align}
  \tilde{C_u}(z)=\frac{C_u(0)}{\tilde{\gamma}(z)/M+z}=\frac{k_\textrm{B}T}{\tilde{\gamma}(z)+Mz}  ,
\end{align}
with the initial condition  $C_u(0)=k_\textrm{B}T/M$, i.e., equipartition of energy, where $z$ is the Laplace variable, $k_B$ is Boltzmann's constant, and $T$ the temperature.
For $C_u(-t)=C_u(t)=C_u^*(t)$, the correlation function $\tilde C_u(z)$  is related with the Fourier transform
\begin{align}
\hat C_u(\omega) =\int_{-\infty}^\infty dt~  e^{i \omega t} C_u(t)
\end{align}
of $C_u(t)$
according to $\hat C_u(\omega) = 2 \textrm{Re} \{\tilde C_u(z=-i\omega) \}$. We will work further in Fourier space for convenience. The velocity autocorrelation function is then given by
\begin{align} \label{Eq:Velocity_autocorrelation_Fourier_Space}
  \hat{C}_u(\omega)=2 \textrm{Re} \left\{ \frac{k_\textrm{B}T}{\hat{\gamma}(\omega)-i \omega M}\right\}.
\end{align}
The explicit expression for $\hat \gamma(\omega)$ is provided in App.~\ref{sec:FrequencyDependentHydrodynamicFrictionTranslation}.

Similarly, we can determine the force autocorrelation function of the random forces (FACF)
\begin{align}
  C_K(t)=\frac{1}{3} \langle \bm{K}(t) \cdot \bm{K}(0)\rangle.
\end{align}
In order to determine $C_K(t)$, we establish a relation between $\langle \hat {\bm u}(\omega_1) \cdot \hat{\bm u}^*(\omega_2)\rangle$
and $\langle \hat{\bm K}(\omega_1) \cdot \hat{\bm K}^*(\omega_2)\rangle$ by Fourier transforming  Eq.~(\ref{Eq_motion_Brownian_general}).
This yields the fluctuation-dissipation relation
\begin{align} \label{Eq:C_K(omega)}
  \hat{C}_K(\omega)=2 k_\textrm{B} T \textrm{Re} \{ \hat{\gamma}(\omega) \}.
\end{align}
The time integral of the force autocorrelation function is related with the zero-frequency friction coefficient according to
\begin{align} \label{Eq:IntegrationFACFequalsGAMMA}
  \int_0^\infty dt ~ C_K(t)=\frac{1}{2} \hat{C}_K(\omega=0)=k_\textrm{B} T \hat{\gamma}(\omega=0).
\end{align}
On the other hand, the time integral of the velocity autocorrelation function, which is by definition the diffusion coefficient $D$, yields
\begin{align} \label{Eq:IntegrationVACFequalsMOBILITY}
  D=\int_0^\infty dt~  C_u(t)=\frac{1}{2} \hat{C}_u(\omega=0)=\frac{k_\textrm{B} T}{\hat{\gamma}(\omega=0)},
\end{align}
which is known as Einstein-Sutherland relation.

Similarly, the equation of motion for the rotational motion is
\begin{align}
I \frac{d {\bm \Omega}}{dt}=-\int_0^t dt'~ \xi(t-t') {\bm \Omega}(t')+{\bm N}(t),
\end{align}
where $I$ is the moment of inertia, $\bm{\Omega}$ is the angular velocity, $\xi(t)$ is the rotational friction, and $\bm{N}$ is the random torque.
Since this equation is mathematically equivalent to Eq. (\ref{Eq_motion_Brownian_general}), all results for the translational motion apply here as well.
In particular, Eqs.~(\ref{Eq:IntegrodifferentialEquation_for_C_u})-(\ref{Eq:IntegrationVACFequalsMOBILITY}) hold, when $M$ is replaced by $I$, $\bm{u}$ by $\bm{\Omega}$, $\gamma$ by $\xi$,
$\bm{K}$ by $\bm{N}$, and $D$ by $D_R$, with the rotational diffusion coefficient $D_R$.

\subsection{Hydrodynamic and local friction}

On long time scales, the MPC solvent can be described by hydrodynamics, whereas on short time scales the molecular chaos assumption applies \cite{ripo:05,huan:12}. For the latter, the relevant time range depends on the collision time $h$.
We will denote the resulting friction coefficients on these time scales as hydrodynamic friction $\gamma_h$ and local friction $\gamma_l$.
In our simulations, we determine the friction coefficients via velocity and force autocorrelation functions.
Theoretical predictions for $C_u(t)$ and $C_K(t)$ can be found by numerical Fourier transformation of Eqs.~(\ref{Eq:Velocity_autocorrelation_Fourier_Space}) and (\ref{Eq:C_K(omega)}),
with  $\hat{\gamma}(\omega)$ given in App.~\ref{sec:FrequencyDependentHydrodynamicFrictionTranslation}.\\
In an analogous manner, $C_\Omega(t)$ and $C_{N}(t)$ are determined by Fourier transformation with $\hat{\xi}(\omega)$ of App.~\ref{sec:FrequencyDependentHydrodynamicFrictionRotation}.\\

\paragraph*{Hydrodynamic friction}

The classical result for the frequency dependent hydrodynamic friction $\hat{\gamma}(\omega)$  (Ref. \cite{chow:73}) assumes local angular momentum conservation for the solvent, i.e., $\eta_1 =\eta_2$ in Eq. (\ref{StressTensorGen}).
The respective derivation of the friction coefficients for translational and rotational motion for the case $\eta_1 \neq \eta_2$, which applies for MPC-SRD-a, is presented in App.~\ref{sec:FrequencyDependentHydrodynamicFrictionTranslation}.
In any case, on long time scales, where the MPC fluid is described by hydrodynamics, the correlation function $C_u(t)$  exhibits the well-known algebraic long-time tail \cite{alde:70,ripo:05,pobl:14}
\begin{align} \label{Eq:Into_LTT}
   C_u(t) \xrightarrow{t \to \infty} \frac{2 k_\textrm{B} T}{3 \rho_0} \left(4 \pi \frac{\eta_2}{\rho_0} |t| \right)^{-3/2}.
\end{align}
Here, $\rho_0$ is the equilibrium mass density.
For a no-slip colloid, $\gamma_h$ is unaffected by angular momentum conservation and reads \cite{yang:15}
\begin{align} \label{eq:stokes_friction_noslip}
  \gamma_h = 6 \pi \eta R .
\end{align}
In contrast, for a colloid with a slip boundary condition, the classical result $\gamma_h = 4 \pi \eta_2 R$ is modified, and the friction coefficient becomes
\begin{align} \label{Eq:ReducedFrictionMain}
  \gamma_h = 6\pi \eta_2 R \frac{\eta_1+\eta_2}{\eta_1+2 \eta_2}
\end{align}
for  $\eta_1 \neq \eta_2$
(cf. Ref. \cite{yang:15} and Eq.~(\ref{Eq:Reduced_friction_omega_zero})).
For MPC-SRD-a, Eq.~(\ref{Eq:ReducedFrictionMain}) reduces to  $\gamma_h \approx 4 \pi \eta R$ for $\eta^c \ll \eta^k$, and to $\gamma_h \approx 3 \pi \eta R$ in the typical case $\eta^c \gg \eta^k$.
In contrast, the stress tensor is symmetric for MPC-SRD+a, i.e.,  $\eta_1=\eta_2$, and the classical result $\gamma_h = 4 \pi \eta R$ is recovered.
The zero-frequency limit of the rotational friction coefficient is
\begin{align}
\xi_h&=(1-\Gamma) 8\pi \eta R^3.
\end{align}
Note that for slip colloids $\Gamma=1$ so that no hydrodynamic drag torque opposes the colloid's rotation.\\

\paragraph*{Local friction}
At short times, the molecular chaos assumption applies for the MPC fluid \cite{ripo:05,huan:12}, which yields $C_K(t)=0 ~ \forall t>0$.
In the time continuum this implies that $C_K(t)$ is proportional to a delta distribution and therefore $\hat{C}_K(\omega)=\hat{C}_K(\omega=0)$.
By means of Eq.~(\ref{Eq:C_K(omega)}), we find $\hat{\gamma}(\omega)=\hat{\gamma}(\omega=0) \equiv \gamma_l$  and
the typical Langevin equation is obtained, which yields
\begin{align} \label{Eq:Cu_exp_decay}
C_u(t)=C_u(0)\exp{(-\gamma_l t/M)}.
\end{align}
We expect to observe this behavior in MPC simulations only approximately and for very short times.
Since MPC is a discrete-time random process and $C_K(t)=0 ~ \forall t>0$, Eq. (\ref{Eq:IntegrationFACFequalsGAMMA}) yields
\begin{align} \label{Eq:gammaSproptoCK}
  \gamma_l=\frac{h}{2 k_B T} C_K(0).
\end{align}
This equation provides a simple relation to measure the local friction coefficient $\gamma_l$ \cite{impe:11}.

In order to find an analytical expression for $\gamma_l$, we have to evaluate various contributions to $C_K(0)$.
The random force $\bm{K}=\bm{K}^k+\bm{K}^c$ comprises the contributions $\bm{K}^k$  due to collisions with MPC particles in the streaming step
and $\bm{K}^c$ due to the change of the ghost-particle momenta during the collision step \cite{impe:11}.
Hence, the autocorrelation functions $\langle \bm{K}^k \cdot \bm{K}^k \rangle$ and $\langle \bm{K}^c \cdot \bm{K}^c \rangle$, as well as the cross-correlation function $\langle \bm{K}^k \cdot \bm{K}^c \rangle$ contribute to $C_K(0)$.
We denote the friction by the correlation $\langle \bm{K}^k \cdot \bm{K}^k \rangle$ as Enskog friction $\gamma_E$.
When no ghost particles are present, as is the case for slip boundary conditions, $\bm{K}^c=0$ and $\gamma_l=\gamma_E$.
A derivation for $\gamma_E$ is presented in App.~\ref{sec:EnskogFriction}, which yields the result previously established in Ref. \cite{whit:10}
\begin{align}
   \gamma_E=\frac{8}{3} \sqrt{2 \pi k_\textrm{B} T \mu} n R^2 \frac{1+(2-\Gamma)\chi M/\mu}{1+\chi M/\mu}.
\end{align}
Here, $n=\rho_0/m$ is the particle density.
We could not find analytical expressions for $\langle \bm{K}^c \cdot \bm{K}^c \rangle$ and $\langle \bm{K}^c \cdot \bm{K}^k \rangle$.
Therefore,  we measure $\gamma_l$ according to Eq.~(\ref{Eq:gammaSproptoCK}) in presence of ghost particles.

For the rotational motion, the treatment is analogous. The local and Enskog friction are (cf. App.~\ref{sec:EnskogFriction})
\begin{align}
  \xi_l&=\frac{h}{2 k_B T} C_N(0), \\
  \xi_E&=(1-\Gamma)\frac{8}{3}\sqrt{2 \pi k_B T \mu} n R^4 \frac{M \chi}{\mu+M \chi}.
\end{align}

\section{Simulations} \label{sec:Simulations}
MPC simulations are performed with the rotation angle $\alpha=130^\circ$ and the mean
number of particles per collision cell $\left< N_c \right>=10$, which corresponds to the equilibrium density $\rho_0=\left< N_c \right>m/a^3$.
We focus on the liquid-like regime of the MPC fluid \cite{ripo:05} by choosing the collision time as $h/\sqrt{ma^2/(k_B T)}=0.05$, which corresponds to a Schmidt number of approximately $100$.
We employ a cubic simulation box of length $L/a=100$ with periodic boundary conditions if not otherwise stated.
Since simulation results shall be compared to hydrodynamic theory, we require values of high accuracy for the viscosities $\eta^k$ and $\eta^c$ for our applied methods (MPC-SRD$\pm$a) and parameters. Measuring viscosities by nonequilibrium simulations using shear flow \cite{wink:09,thee:15,huan:15} yields $\eta=7.45\sqrt{m k_B T}/a^2$ and $\eta^V=5.40\sqrt{m k_B T}/a^2$ for MPC-SRD+a, as well as $\eta^k=0.3 \sqrt{m k_B T}/a^2$ and $\eta^c=16.2 \sqrt{m k_B T}/a^2$ for MPC-SRD-a.
In the following, we assume a neutrally buoyant colloid, i.e., $M=(4\pi/3)\rho_0 R^3$.

On the order of $100$ independent simulations of $10^6$ time steps each were performed for a given parameter set to extract autocorrelation functions.\\

\subsection{Autocorrelation functions}
\begin{figure} 
\includegraphics*[width=\columnwidth]{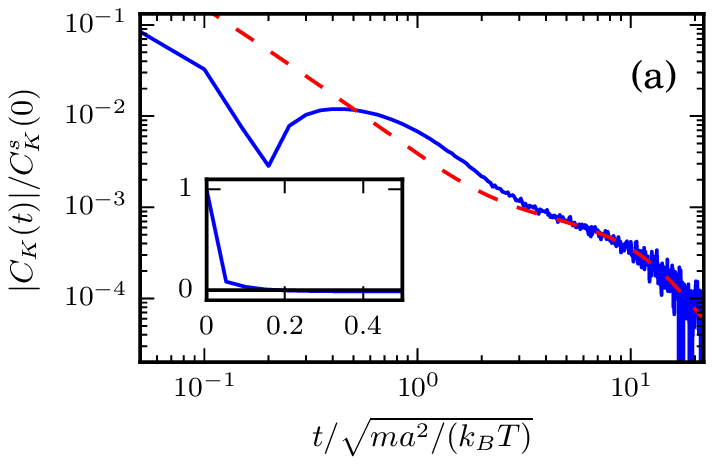}
\includegraphics*[width=\columnwidth]{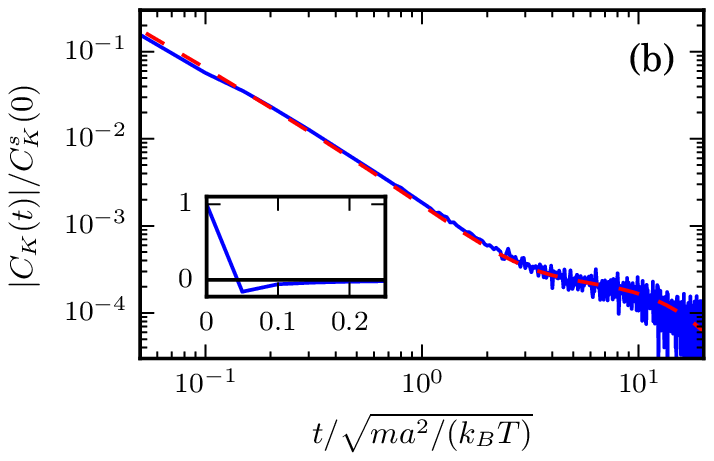}
\caption{\label{fig:FACF_SLIP=1_AMC=0_h=0.05_R=6_L=100} 
  (Color online) (a) Force autocorrelation function (FACF)  of a slip colloid in a MPC-SRD-a  fluid (blue solid line).
  The inset shows the FACF on a linear scale; the FACF becomes negative at $t=0.2 \sqrt{ma^2/(k_B T)}$.
  (b) FACF of a no-slip colloid in a MPC-SRD+a fluid. In both cases, the radius of the colloid is $R/a=6$.
  Note that the prediction by hydrodynamic theory (red, dashed line) diverges for $t \to 0$. Therefore, both curves are normalized by the simulation value for $t=0$, denoted as $C_K^s(0)$.
}
\end{figure}

\begin{figure} 
\includegraphics*[width=\columnwidth]{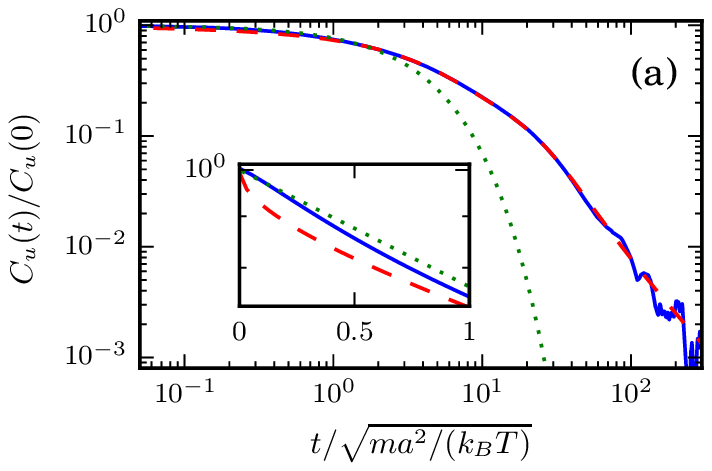}
\includegraphics*[width=\columnwidth]{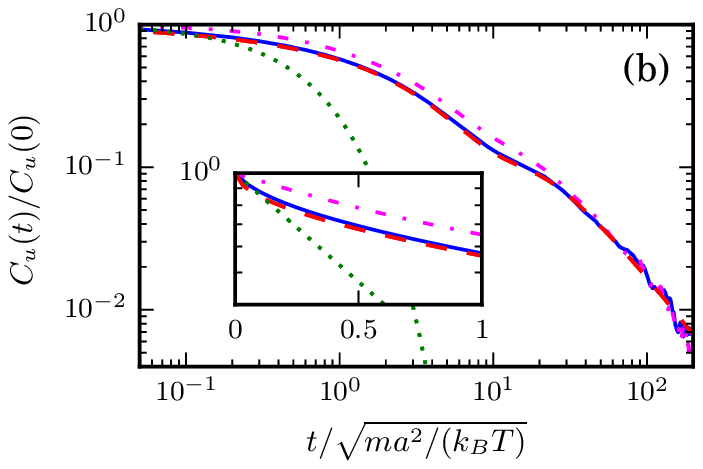}
\caption{\label{fig:VACF_SLIP=1_AMC=0_h=0.05_R=6}
  (Color online) (a) VACF of a colloid with slip boundary conditions in a MPC-SRD-a fluid, and (b) with no-slip boundary conditions in a MPC-SRD+a fluid.
  The solid line (blue) corresponds to the simulation results,  the dashed line (red) is the prediction by hydrodynamic theory, and the dotted line (green) is calculated according to 
  Eq. (\ref{Eq:Cu_exp_decay}).
  The dashed-dotted line (magenta) in (b) is the simulation result without ghost particles.
  The insets display the first few time steps in a semi-logarithmic representation.
  In both cases, the radius of the colloid is $R/a=6$.
}
\end{figure}

\begin{figure} 
\includegraphics*[width=\columnwidth]{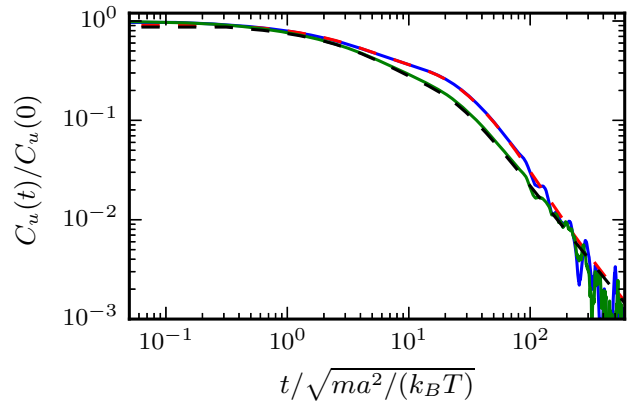}
\caption{\label{fig:VACF_AMC=0_h=0.12_AMC=1_h=0.05}
  (Color online) VACFs of colloids with slip boundary conditions in a MPC-SRD-a and a MPC-SRD+a fluid. The upper curves (red and blue )correspond to MPC-SRD-a with $h/\sqrt{ma^2/(k_\textrm{B} T)}=0.12$ and the lower ones (black and green) to MPC-SRD+a with $h/\sqrt{ma^2/(k_\textrm{B} T)}=0.05$.
  The colloid radius is  $R/a=6$ and the viscosity $\eta=7.45 \sqrt{m k_\textrm{B} T}/a^2$.
  The theoretical curves (dashed) are calculated by means of Eqs. (\ref{Eq:Velocity_autocorrelation_Fourier_Space}) and (\ref{Res_gamma_of_omega}) with the appropriate viscosities.
}
\end{figure}

\begin{figure}
\includegraphics*[width=\columnwidth]{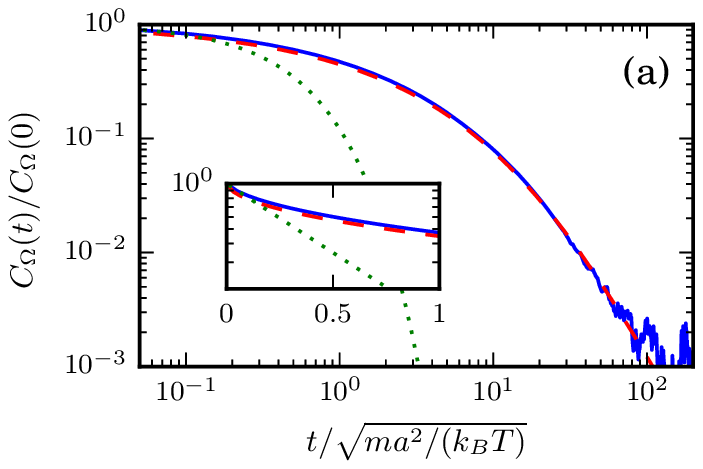}
\includegraphics*[width=\columnwidth]{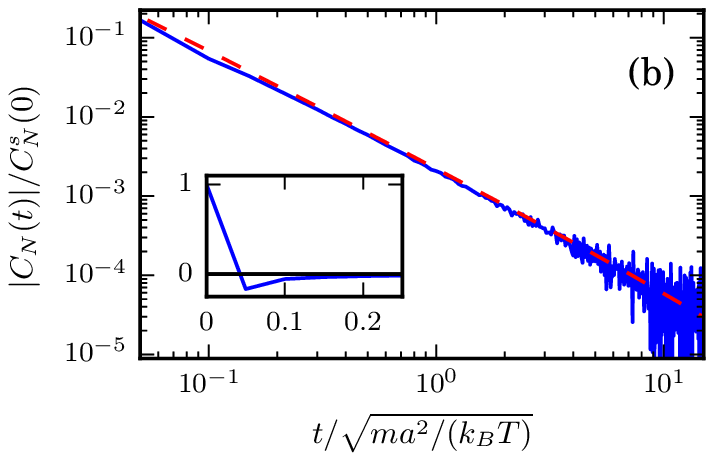}
\caption{\label{fig:AVACF_SLIP=0_AMC=1_h=0.05_R=6}
 (Color online) (a) Angular velocity autocorrelation function and (b) torque autocorrelation function for a colloid with no-slip boundary conditions in a MPC-SRD+a fluid. The colloid radius is $R/a=6$  and the fluid viscosity $\eta=7.45 \sqrt{m k_\textrm{B} T}/a^2$.
 The box sizes are $L=100a$ (a) and $L=60a$ (b), respectively.
  The inset in (a) shows the first few time steps on a semilogarithmic scale; that in  (b) presents the torque autocorrelation function on a linear scale.
  The torque autocorrelation function in (b) reaches its $t^{-5/2}$ long-time tail at $t/\sqrt{ma^2/(k_\textrm{B}T)} \gtrsim 10^2$ which is not visible in the plotted time domain.
  Since the theoretical hydrodynamic-torque autocorrelation function (dashed, red) diverges for $t \to 0$, both curves are normalized by the simulation value at $t=0$, denoted as $C_N^s(0)$.
}
\end{figure}

In simulations, the velocity autocorrelation function is determined by measuring the velocity of a freely diffusing colloid in the MPC solvent.
For the force autocorrelation function, we fix the colloid center at the origin of the reference frame by means of a constraining force ${\bm F}^C(t)$.
The random force follows then as ${\bm K}(t)=-{\bm F}^C(t)$ \cite{lee:04,impe:11}, while ${\bm F}^C$ is calculated as momentum transfer per time step $h$, with contributions due to collisions of MPC particles in the streaming step and  due to the change of the ghost particles' momenta during the collision step.

Figure \ref{fig:FACF_SLIP=1_AMC=0_h=0.05_R=6_L=100} shows two typical force autocorrelation functions.
In case of the slip colloid, the FACF at $t=0$ matches the expected Enskog value $2 k_\textrm{B} T \gamma_E/h$ (cf. Eq. (\ref{Eq:gammaSproptoCK}) or App.~\ref{sec:EnskogFriction}) very well. 
However, for the no-slip colloid, $C_K(0)$ is only captured by Enskog theory if ghost particles are neglected, as can be expected for non-zero correlations  $\langle \bm{K}^c \cdot \bm{K}^c \rangle$.
Taking into account the momentum transfer due to ghost particles, we obtain a significantly larger value of $C_K(0)$, and hence, of the local friction coefficient.

Within the molecular chaos assumption, the FACF is zero for all $t>0$, which implies an exponentially decaying VACF.
In fact, in our MPC simulation, the FACF decreases substantially after one time step, but instead of zero, it assumes about $10\%$ of its initial value for the parameters of Fig. \ref{fig:FACF_SLIP=1_AMC=0_h=0.05_R=6_L=100} (a). This explains the approximate exponential decay of the VACF for the first few time steps, as reported in Refs.~\cite{lee:04, padd:06,whit:10,belu:11}.
As Fig. \ref{fig:FACF_SLIP=1_AMC=0_h=0.05_R=6_L=100} shows, the FACF is rather noisy, and an average over many independent realizations is required to achieve a smooth curve on long time scales.

The corresponding velocity autocorrelation functions are much smoother, as revealed by Fig. \ref{fig:VACF_SLIP=1_AMC=0_h=0.05_R=6}.
At $t=0$, $C_u(t=0)=k_B T/M$, as expected. For short times $t>0$, the simulation data slightly exceed the theoretical prediction until the hydrodynamic regime is reached.
For longer times, we observe the long-time tail (cf. Eq. (\ref{Eq:Into_LTT}). The oscillations visible in Fig. \ref{fig:VACF_SLIP=1_AMC=0_h=0.05_R=6} for long times  originate from sound modes and are a consequence of the finite compressibility of the MPC fluid combined with the periodic boundary conditions \cite{huan:12}. This also leads to an exponential decay of the correlation function on long time scales \cite{huan:12}.
We determine the VACF for colloids of radii $R/a=1,2,\dots,8$, and find that for no-slip boundary conditions the VACF follows the hydrodynamic prediction well for radii $R>2a$,
while for slip colloids, $R \ge 3a$ is required.

A detailed analysis of the VACF at $t=0$ shows a small deviation from the equipartition value $C_u(0)=k_BT/M$, which vanishes with increasing colloid size.
For a slip colloid of radius $R=3a$ in a MPC-SRD+a fluid, $C_u(0)$ is just $1\%$ larger than  the expected value, while for no-slip colloids the theoretical value is exceeded by $5\%$  for  $R=2a$ and by $2\%$ for $R=3a$.
Working with $R/a\geq 3$, these values are acceptable, and we do not see a broadened Maxwell-Boltzmann distribution due to ghost particles as reported in Ref. \cite{whit:10}.
A disregard of ghost particles, as suggested in Ref~\cite{whit:10}, leads to strong deviations between VACF obtained in simulations and from hydrodynamic theory for $t>0$ as shown by Fig. \ref{fig:VACF_SLIP=1_AMC=0_h=0.05_R=6} (b).
Hence, ghost particles are essential to obtain a good representation of no slip boundary conditions.

To elucidate the relevance of angular momentum conservation on the VACF, we perform simulations of a slip colloid in MPC-SRD+a with the collision step $h/\sqrt{ma^2/(k_\textrm{B} T)}=0.05$ and in MPC-SRD-a with   $h/\sqrt{ma^2/(k_\textrm{B} T)}=0.12$, respectively. By this choice, both fluids possess approximately the same shear viscosity $\eta=7.45 \sqrt{m k_\textrm{B} T}/a^2$.
As Fig. \ref{fig:VACF_AMC=0_h=0.12_AMC=1_h=0.05} shows, the velocity autocorrelation functions deviate from each other, but both are well described by hydrodynamic theory
with the appropriate stress tensor. The colloid VACF for the MPC-SRD-a fluid is consistently larger than that for the MPC-SRD+a fluid, which corresponds to a reduced friction of the MPC-SRD-a fluid (cf. Eq. (\ref{Eq:ReducedFrictionMain})).
This reconfirms the result of Ref. \cite{yang:15} that the friction of slip colloids is reduced for non-angular-momentum conserving MPC fluids.


Finally, Fig.~\ref{fig:AVACF_SLIP=0_AMC=1_h=0.05_R=6} displays simulation results for the angular-velocity-autocorrelation function as well as the torque-autocorrelation function for
a no-slip colloid, which both agree well with hydrodynamic theory. Similar to the force autocorrelation, the torque autocorrelation function is very noisy.

\subsection{Diffusion coefficient} \label{sec:IntegralExtrapolation}
\begin{figure} 
\includegraphics*[width=\columnwidth]{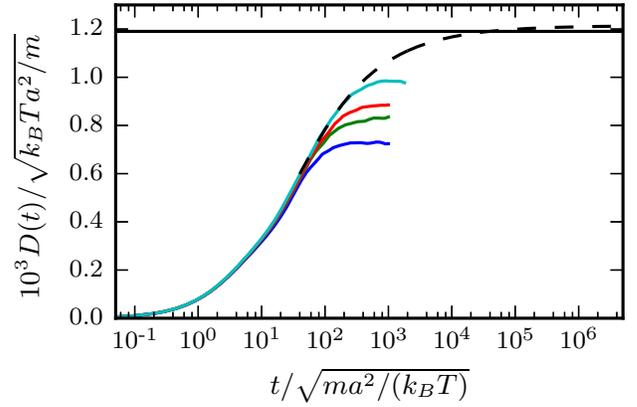}
\caption{
\label{fig:IntegralExtrapolation}
(Color online) Integrals according to Eq.~(\ref{eq:diffusion}) of the VACF for the simulation box sizes $L/a=40$ (blue), $50$ (green),  $60$ (red), $80$ (cyan) (bottom to top).
The dashed line (black) indicates the integration of the combined VACF from simulation up yo $t=t_0 \approx 40 \sqrt{ma^2/(k_B T)}$ and the theoretical expression of hydrodynamics beyond.
The horizontal line (black) marks the diffusion coefficient from hydrodynamic theory.
}
\end{figure}

\begin{figure} 
\includegraphics*[width=\columnwidth]{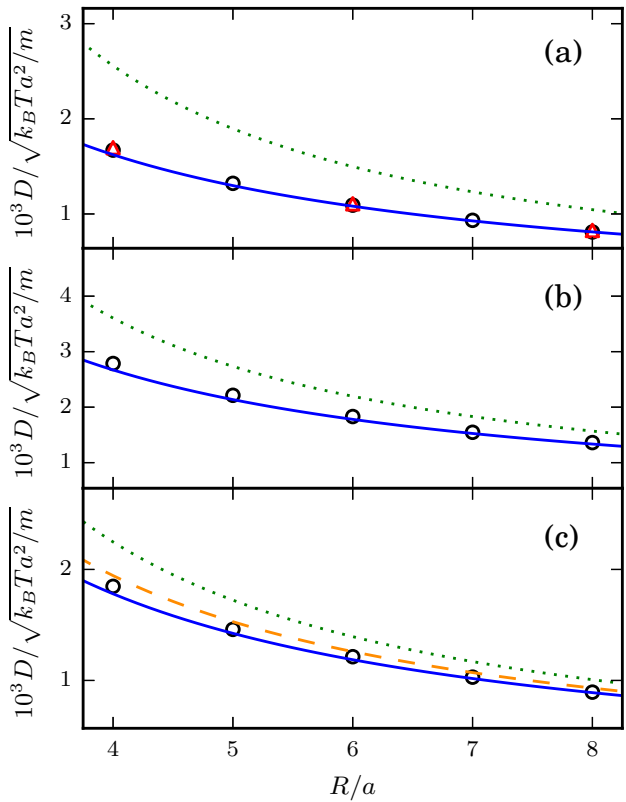}
\caption{\label{fig:D_of_R}
 (Color online) Diffusion coefficients as function of the radius of colloids with slip boundary conditions (a) in a MPC-SRD-a fluid,  (b)  in a MPC-SRD+a fluid, and  (c)  a  colloid with no-slip boundary conditions in a MPC-SRD+a fluid.
 The solid lines (blue) represent $D_h$, the dotted lines (green) $D_h+D_E$, and the dashed lines (orange) $D_h+D_l$.
 Open circles (black) correspond to values extracted from the VACF (cf. sec. \ref{sec:IntegralExtrapolation}), while the triangles (red) are obtained by integration of the FACF.
}
\end{figure}

As is well known, the diffusion coefficient of a particle in a system with periodic boundary conditions is system-size dependent \cite{pier:92,duen:93,yeh:04,padd:06,ladd:09,huan:13}.
In order to find the asymptotic diffusion coefficient for an infinite system ($L\to \infty$), we integrate the simulation data for the VACF from $t=0$ to $t_0>0$ and subsequently integrate the theoretical correlation function from $t=t_0$ to $t \to \infty$ \cite{huan:13,pobl:14}.
Hereby, $t_0$ has been chosen such that hydrodynamic theory applies for $t>t_0$.
This procedure is illustrated in Fig. \ref{fig:IntegralExtrapolation}, where the expression
\begin{align} \label{eq:diffusion}
D(t) = \int_0^t dt'  C_u(t'0)
\end{align}
is displayed for various box sizes, and for a VACF comprised of the numerical results for $L/a=80$ up to $t_0=40 \sqrt{ma^2/(k_BT)}$ and the theoretical expression following from Eq.~(\ref{Eq:Velocity_autocorrelation_Fourier_Space}) for longer times.

In this way, we determine the diffusion coefficients of  slip and no-slip colloids for several radii.
The results are presented in Fig. \ref{fig:D_of_R}. Evidently, the diffusion coefficients are in close agreement with the prediction by hydrodynamics at large colloid radii.
However, for small radii, we observe certain deviations, which we attribute to the effect of local friction.
In general, the simulation data are by far closer to the hydrodynamic diffusion coefficient $D_h$ than to the combination with the Enskog expression, i.e., $D_h+D_E$, where $D_h=k_B T/\gamma_h$ and $D_E=k_B T/\gamma_E$.

In case of no-slip colloids, we have to compare $D_h$ to $D_h+D_l$, where $D_l=k_B T/\gamma_l$ accounts for local interactions. Note that  $D_l$ is small compared to $D_h$,  because $\gamma_l$ is very large.
Our simulation results are about midway in-between the predictions $D_h$ and $D_h+D_l$. As a consequence,
for both, slip and no-slip colloids, hydrodynamics dominates at large $R$ and the simulation results are well described by $D_h$.

\subsection{Colloid sedimentation and diffusion coefficient}
\begin{figure} 
\includegraphics*[width=\columnwidth]{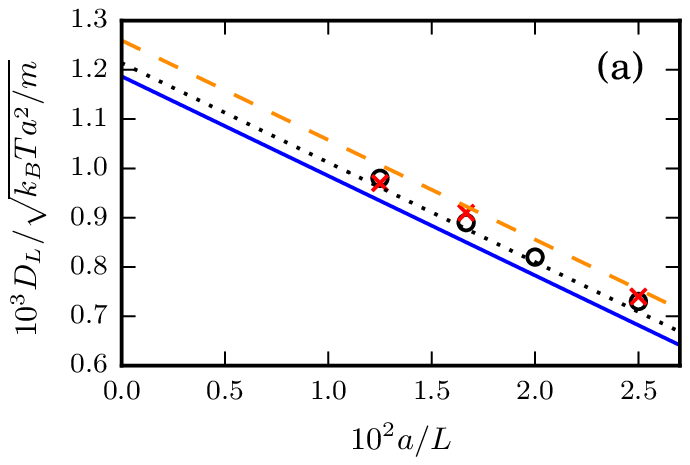}
\includegraphics*[width=\columnwidth]{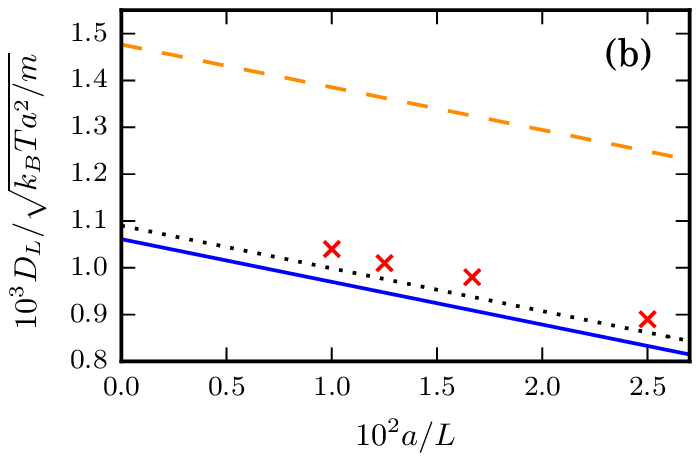}
\caption{\label{fig:DuenwegExtrapolation}
  (Color online) Finite-system-size diffusion coefficients  $D_L$  of colloids with (a) no-slip boundary conditions in a MPC-SRD+a  and (b) slip boundary conditions in a MPC-SRD-a fluid. The colloid radius is  $R/a=6$.
  The open circles correspond to the  plateau values of Fig. \ref{fig:IntegralExtrapolation}. The values indicated by crosses (red) are obtained by sedimentation simulations.
  The lines follow from  Eq. (\ref{Eq:DuenwegExtrapolation}), where we insert $D=D_h+D_l$ for the dashed orange line,
  $D=D_h$ for the blue solid line, and we substitute $D$ by the value obtained from the integral extrapolation approach of section \ref{sec:IntegralExtrapolation} for the black dotted line.
}
\end{figure}

We can extract finite-system-size diffusion coefficients $D_L$ by the plateau values in Fig. \ref{fig:IntegralExtrapolation}.
Complementary, we perform simulations of sedimenting colloids by  constantly accelerating a colloid with the acceleration $g=10^{-3} k_B T/(m a)$.
At the same time, the fluid experiences the acceleration $g_f=-g/(\rho_0 L^3/M -1)$, such that the center of mass of the total system remains at rest.
For the selected values, the Reynolds number is approximately $0.1$.
When a steady state velocity $u$ is reached, we obtain the finite-system-size friction coefficient as $\gamma_L=Mg/u$, and the diffusion coefficient as $D_L=k_B T/\gamma_L$ (Einstein-Sutherland relation).

For a hydrodynamic solvent, a relation between $D_L$ and the diffusion coefficient of an infinite system $D$ has been provided in Ref.~\cite{duen:93} to first order in $R/L$,
\begin{align} \label{Eq:DuenwegExtrapolation}
  D_L=D-\frac{2.837 k_\textrm{B} T}{6 \pi \eta L}.
\end{align}
Since this expression is derived for point particles, the finite-size correction should hold for both slip and no-slip colloids.
We consider colloids of radius  $R/a=6$, for which hydrodynamics should dominate in our simulations, and thus, Eq.~(\ref{Eq:DuenwegExtrapolation}) should apply.\\
Results for $D_L$ are presented in Fig.~\ref{fig:DuenwegExtrapolation} together with the theoretical expression Eq.~(\ref{Eq:DuenwegExtrapolation}) for various systems sizes $L$.
Our values obtained by sedimentation are only about $2\%$ higher than those determined via the VACF, which is an excellent agreement taking the accuracy of our methods into account.

We return to the question whether diffusivity is additive, i.e., $D=D_h+D_l$.
According to the sedimentation data for no-slip colloids, $D=D_h+D_l$ seems to be a decent approximation, with an error of about $2\%$.
For slip colloids, however, the prediction $D=D_h+D_l$ exceeds the measured values by about $30\%$.
Hence, for slip colloids the diffusivities are not additive in MPC.

\section{Summary and conclusions} \label{sec:Conclusions}

We have determined velocity and force autocorrelation functions of rigid spherical colloids dispersed in a MPC fluid. Both, slip and no-slip boundary conditions on the colloid surface have been considered.  For slip boundary conditions, MPC fluids with (MPC-SRD+a) and without (MPC-SRD-a) angular-momentum conservation have been employed. In contrast, for no-slip boundary conditions only a MPC-SRD+a fluid has been utilized, because the coupling between the colloidal rotational degrees of freedom and the fluid requires a proper angular momentum transfer.
We have verified the reduction of hydrodynamic friction of colloids with slip boundary conditions in a MPC-SRD-a fluid compared to the respective Stokes law. As derived in Ref. \cite{yang:15}, the friction coefficient of such a colloid  is given by an expression (Eq. (\ref{Eq:Reduced_friction_omega_zero})),
which reduces to $\gamma_h \approx 3 \pi \eta R$ for $\eta^c \gg \eta^k$, in contrast to the classical result $\gamma_h=4 \pi \eta R$ for an angular-momentum-conserving fluid. We like to stress that differences in the transport coefficients of angular- and non-angular-momentum-conserving fluids are a general problem of fluid simulations and not particular to MPC.

By measuring correlation functions for no-slip colloids, we found that ghost fluid particles inside the colloid are essential for a proper hydrodynamic colloid-solvent coupling. We did not observe a broadening of the Maxwell-Boltzmann distribution due to ghost particles, as discussed in Ref. \cite{whit:10}, i.e., a notably deviation of  $C_u(0)$ from $k_B T/M$.
Hence, ghost particles are essential to properly describe the dynamics of colloids with no-slip boundary conditions.

In addition, we have extracted diffusion coefficients from correlation functions and indirectly via colloid sedimentation. Most importantly, our simulations clearly show that the colloid diffusion coefficient is dominated by hydrodynamics in the parameter regime where MPC is liquid-like, i.e., for $\eta^c \gg \eta^k$. Local friction yields only a minor contribution to the overall diffusion coefficient, and disappears with increasing colloid radius. Thereby, for no-slip colloids, the sum of the  local and hydrodynamic diffusion coefficients is an acceptable approximation for the total diffusivity. This sum slightly overestimates the simulation result, but the deviation disappears with increasing colloid radius, and the  Stokes-Einstein relation is satisfied. We have confirmed the result for the time steps $h/\sqrt{ma^2/(k_BT)}=0.1, 0.02$. Thereby, all considered collision-time steps are well within the liquid regime of MPC-SRD \cite{ripo:05}.
However, for colloids with slip boundary conditions, the total diffusion coefficient is not given by the sum of the hydrodynamic and local diffusion coefficients, regardless of angular-momentum conservation. The combination of both significantly overestimates the diffusivity. As for other colloids, the simulation result is well described by the hydrodynamic diffusivity.

In the literature, colloids with slip boundary conditions have also been modeled by finite-range (steep) central potentials between colloids and MPC fluid particles \cite{lee:04,padd:06}. 
For infinitely steep interaction potentials, this approach should be equivalent to the method of specular reflection applied in this paper. For a finite interaction range, the interpretation of simulation results is less straight-forward, since an effective hydrodynamic
radius has to be introduced as a fit parameter. But also in this case, we expect the non-additivity of diffusion coefficients to prevail.

Our studies clearly underline the dominance of hydrodynamic interactions in colloid diffusion. Even for moderately large colloids, hydrodynamics dominates already. The necessity of a local friction contribution in previous studies for colloids with slip boundary conditions may partially originate from an overestimation of the hydrodynamic friction coefficient by assuming the applicability of the Stokes law. The actual coefficient is up to a factor of 3/4 smaller, and hence the diffusion coefficient correspondingly larger.

\begin{acknowledgments}
We thank Adam Wysocki, Anoop Varghese, and B.U. Felderhof for helpful discussions.
\end{acknowledgments}

\appendix

\section{Frequency-dependent hydrodynamic friction} \label{sec:FrequencyDependentHydrodynamicFriction}
\subsection{Translational friction} \label{sec:FrequencyDependentHydrodynamicFrictionTranslation}
In the following, we derive expressions for the friction coefficient $\hat{\gamma}(\omega)$ of a spherical colloid with slip or no-slip boundary conditions moving with
velocity $\bm{u}(t)=u(t)(0,0,1)^T$ in a fluid, with a stress tensor specified by Eq. (\ref{StressTensorGen}).
Thereby, we follow the derivation of Ref. \cite{chow:73}.
The Navier-Stokes equation
\begin{align}
 \rho \left(\frac{\partial {\bm v}}{\partial t} + {\bm v} \cdot ({\bm \nabla} {\bm v}^T) \right)= {\bm \nabla} \cdot {\bm \sigma} ,
\end{align}
together with the continuity equation, can be linearized in the velocity $\bm{ v}(\bm{r},t)$, the pressure, and density fluctuations \cite{boon:80},
which yields in frequency space
\begin{align}
 -i \omega \rho_0 \hat {\bm v}&=- {\bm \nabla} \hat{p} +\eta_2 \Delta \hat{\bm v} +(\eta_1+\eta_3) {\bm \nabla} ({\bm \nabla} \cdot \hat{\bm{v}}), \label{NavSto_freq}\\
 i \omega \hat{\rho} &=\rho_0 ({\bm \nabla} \cdot \hat{\bm{v}}), \label{Cont_freq}\\
 \hat{p}&=\hat{\rho} c^2. \label{EqState_freq}
\end{align}
Here, $\rho_0$ is the equilibrium mass density.
Equation (\ref{EqState_freq}) is the linearized relation between pressure and density, i.e., $c^2=\partial p/ \partial \rho$, where  $c$ is either the adiabatic or  isothermal sound velocity.
MPC obeys the ideal-gas equation of state \cite{wink:09}, and in simulations we apply the MBS-thermostat, hence $c=\sqrt{k_B T /m}$.
The pressure can be eliminated from Eq. (\ref{NavSto_freq}) by means of Eqs.~(\ref{Cont_freq}) and (\ref{EqState_freq}), which yields
\begin{align}
 -i \omega \rho_0 \hat{\bm{v}}=-\eta_2 {\bm \nabla}\times({\bm \nabla} \times \hat{\bm{v}})+\frac{i \omega \rho_0}{\beta^2} {\bm \nabla} ({\bm \nabla} \cdot \hat{\bm{v}}),
 \label{NavSto_freq_elim}
\end{align}
where we define
\begin{align}
 \beta^2=\omega^2 \left( c^2-\frac{i \omega (\eta_1+\eta_2+\eta_3)}{\rho_0}\right)^{-1}, \ \textrm{Im}\{ \beta \}>0.
\end{align}
With the Helmholtz decomposition
\begin{align}
 \hat{\bm{v}}={\bm \nabla} \phi + {\bm \nabla} \times \bm{ A},
\end{align}
Eq. (\ref{NavSto_freq_elim}) becomes
\begin{align}
 \Delta \phi + \beta^2 \phi&=0, \\
 {\bm \nabla}\times({\bm \nabla} \times \bm{ A}) &= \alpha^2 \bm{ A} ,
\end{align}
with $\alpha^2=i \omega \rho_0/\eta_2, \ \textrm{Im}\{ \alpha \}>0$.
In terms of spherical coordinates $(r,\varphi,\theta)$, the Ansatz \footnote{Note that $\phi$ is a scalar that should depend linearly on $\hat{\bm{u}}$. Hence, we assume $\phi$ to be proportional to the scalar product
of $\hat{\bm{u}}$ and $\bm{r}$, i.e., $\phi=\hat{u} \cos (\theta) h(r)$.
The vector $\bm{A}$ should also be linear in $\hat{\bm{u}}$, and since $\hat{\bm{v}}$ is a polar vector, $\bm{A}$ has to be an axial vector.
This motivates the cross product $\bm{A}=\hat{\bm{u}} \times {\bm \nabla} f(r)$. \cite{land:59} }
$\phi=\hat{u} \cos (\theta) h(r)$ and $\bm{ A}=\hat{\bm{u}} \times {\bm \nabla} f(r)$,
yields ordinary differential equations
for $h(r)$ and $f(r)$, with the solutions
\begin{align}
 h(r)&=(r^{-2}-i\beta r^{-1})e^{i \beta r}, \\
 f(r)&=r^{-1} e^{i \alpha r}.
\end{align}
We construct a solution as the linear combination
\begin{align}
 \hat{\bm{v}}=c_2 {\bm \nabla} \phi -c_1 {\bm \nabla} \times \bm{ A}.
\end{align}
The coefficients $c_1$ and $c_2$ are determined by slip boundary conditions,
\begin{align}
 \hat{v}_r(r=R)&=\hat{u} \cos(\theta), \label{Eq:Hydro_Slip_Boundary1} \\
 \hat{\sigma}_{\theta r}(r=R)&=0. \label{Eq:Hydro_Slip_Boundary2}
\end{align}
or no-slip boundary conditions,
\begin{align}
 \hat{\bm{v}}(r=R)=\hat{\bm{u}}. \label{Eq:Hydro_NoSlip_Boundary}
\end{align}
Note that for Eqs. (\ref{Eq:Hydro_Slip_Boundary1})-(\ref{Eq:Hydro_NoSlip_Boundary}), we chose a reference frame, in which the origin always lies in the colloid's center \cite{haug:73}.
The force $\hat{\bm{F}}(\omega)$  on the colloid can be evaluated by a surface integral \cite{land:59}
\begin{align} \label{ForceEqualsIntSigma}
\begin{split}
 \hat{\bm{F}}&= \int_{r=R} d^2r ~\hat{\bm{\sigma}} ~{\bm r}/r \\
 &= \int_{r=R} d^2r \left( -\sin (\theta) \hat{\sigma}_{\theta r} + \cos (\theta) \hat{\sigma}_{r r}\right) (0,0,1)^T ,
\end{split}
\end{align}
where $\hat{\sigma}_{\theta r}$ and $\hat{\sigma}_{r r}$ follow by Eqs. (\ref{StressTensorGen}), (\ref{Cont_freq}), and (\ref{EqState_freq})
\begin{align}
 \hat{\sigma}_{\theta r}&=\eta_1 \left( \frac{1}{r} \frac{\partial \hat{v}_r}{\partial \theta} -\frac{\hat{v}_\theta}{r} \right) + \eta_2 \frac{\partial \hat{v}_\theta}{\partial r}, \\
 \hat{\sigma}_{r r}&=\left(\eta_3 - \frac{\rho_0 c^2}{i \omega} \right) {\bm \nabla} \cdot \hat{\bm{v}} + (\eta_1+\eta_2) \frac{\partial \hat{v}_r}{\partial r}.
\end{align}
With the definition of $\hat{\gamma}(\omega)$ by $ \hat{\bm{F}}=- \hat{\gamma}(\omega) \hat{\bm{u}}$, Eq. (\ref{ForceEqualsIntSigma}) yields
\begin{align}
 \hat{\gamma}(\omega)=-\frac{4 \pi}{3} \eta_2 R x^2 \left[ Q(1-y)+2P(x-1)\right]  , \label{Res_gamma_of_omega}
\end{align}
with the abbreviations
\begin{align}
  x=i \alpha R, \hspace{1 ex} y=i \beta R, \hspace{1 ex} P=\frac{c_1}{R^3} e^{i \alpha R}, \hspace{1 ex} Q=\frac{c_2}{R^3} e^{i \beta R}.
\end{align}
\paragraph*{Slip boundaries:}
Determining the coefficients $c_1$ and $c_2$ by the slip boundary conditions (\ref{Eq:Hydro_Slip_Boundary1}) and (\ref{Eq:Hydro_Slip_Boundary2}), we find
\begin{align}
\begin{split}
 P&=(\eta_1+\eta_2)(y^2-3y+3)/\Delta , \\
 Q&=\left[\eta_1(-3+3x-x^2)+\eta_2(-3+3x-2x^2+x^3) \right]/\Delta , \\
 \Delta&=(\eta_2 x^3-(\eta_1+2\eta_2)x^2)(-2+2y-y^2) \\
      & \quad +(\eta_1+\eta_2)(1-x)y^2.
\end{split}
\end{align}
From Eqs.~(\ref{Eq:C_K(omega)}) and(\ref{Res_gamma_of_omega}), and the analog of Watson's lemma for Fourier transformations \cite{olve:10},
we find the asymptotic long-time behavior of the random-force autocorrelation function
\begin{align} \label{FACF_LongTimeTail}
 \frac{C_K(t)}{k_\textrm{B} T}\xrightarrow{t \to \infty}
 -3 \pi \eta_2 R^2 \frac{(\eta_1+\eta_2)^2}{(\eta_1+2\eta_2)^2} \sqrt{\frac{\rho_0}{\pi \eta_2}} |t|^{-3/2}.
\end{align}
Similarly, the long-time tail of the velocity autocorrelation function follows as \cite{haug:73}
\begin{align}
   \frac{C_u(t)}{k_\textrm{B} T}\xrightarrow{t \to \infty} \frac{2}{3 \rho_0} \left(4 \pi \frac{\eta_2}{\rho_0} |t| \right)^{-3/2}.
\end{align}
In the Stokes limit $\beta \to 0, \ \alpha \to 0$, we obtain the friction coefficient  \cite{yang:15}
\begin{align} \label{Eq:Reduced_friction_omega_zero}
 \hat{\gamma}(\omega=0)=6\pi \eta_2 R \frac{\eta_1+\eta_2}{\eta_1+2 \eta_2} .
\end{align}

\paragraph*{No-slip boundaries:}
Determining the coefficients $c_1$ and $c_2$ by the no-slip boundary conditions (\ref{Eq:Hydro_NoSlip_Boundary}), we find
\begin{align}
\begin{split}
 P&=(y^2-3y+3)/\Delta, \\
 Q&=-(x^2-3x+3)/\Delta, \\
 \Delta&=x^2y^2-y^2x-2x^2y+y^2+2x^2 ,
\end{split}
\end{align}
and for the long-time tail of the random-force autocorrelation function
\begin{align} \label{FACF_LongTimeTail_noslip}
 \frac{C_K(t)}{k_\textrm{B} T}\xrightarrow{t \to \infty}
 -3 \pi \eta_2 R^2 \sqrt{ \frac{\rho_0}{\pi \eta_2}} |t|^{-3/2},
\end{align}
while the long-time tail of the velocity autocorrelation function is the same as for slip boundary conditions.
The friction coefficient in the Stokes limit is $\hat{\gamma}(\omega=0)=6 \pi \eta R$.

\subsection{Rotational friction} \label{sec:FrequencyDependentHydrodynamicFrictionRotation}

For a  rotating sphere with no-slip boundary conditions  $\hat{\bm{v}}(r=R)=\hat{\bm{\Omega}} \times \bm{r}$.
Since the rotational motion of the no-slip colloid should not excite longitudinal sound modes, we can restrict ourselves to an incompressible description of the fluid \cite{jr:77}.
The linearized incompressible Navier-Stokes equation can be solved by the Ansatz
\begin{align}
\hat{\bm{v}}=\bm{\nabla}\times \bm{A},
\end{align}
with $\bm{A}=f(r)\hat{\bm{\Omega}}$ \footnote{Since $\bm{v}$ is a polar vector, $\bm{A}$ has to be an axial vector. But as $\hat{\bm{\Omega}}$ already is
an axial vector, and $\hat{\bm{v}}$ should be linear in $\hat{\bm{\Omega}}$, the Ansatz $\bm{A}=f(r)\hat{\bm{\Omega}}$ is well motivated \cite{land:59}.}.
The torque is found by $\hat{\bm{T}}=\int d^2r~  \bm{r} \times \hat{\bm{\sigma}} ~ \bm{r}/r$,  which yields
\begin{align}
  \hat{\xi}(\omega)=\frac{8\pi \eta_2 R^3}{3} \frac{3-3x+x^2}{1-x},
\end{align}
with  $x=i \alpha R$ and $\alpha^2=i \omega \rho_0/\eta_2, \textrm{Im}\{ \alpha \}>0$.
In the Stokes limit $\omega \to 0$, we obtain the friction coefficient $\xi_h=8 \pi \eta R^3$.

The long-time tail of the angular-velocity autocorrelation and the random-torque autocorrelation function are given by
\begin{align}
  \frac{C_\Omega(t)}{k_\textrm{B} T} &\xrightarrow{t \to \infty}
  \frac{\pi}{\rho_0} \left(4\pi \frac{\eta_2}{\rho_0} |t| \right)^{-5/2}, \\
  \frac{C_N(t)}{k_\textrm{B} T} &\xrightarrow{t \to \infty}
  -2 \sqrt{\frac{\pi \rho_0^3}{\eta_2}} R^6 |t|^{-5/2}.
\end{align}

\section{Enskog friction} \label{sec:EnskogFriction}
For short times, the force autocorrelation function of a colloid is determined by uncorrelated collisions with fluid particles (molecular chaos). Hence,
during the first streaming and collision step, no hydrodynamic correlations build up. (We will neglect ghost particles for the following considerations.)
Since force is change of momentum per time, the force $\bm{K}$ on the colloid during a streaming step is
\begin{align}
 \bm{K}= \frac{1}{h} \sum_{k=1}^N \bm{J}_k .
\end{align}
Hence, the force-autocorrelation function at time $t=0$ is
\begin{align}
\begin{split}
 \langle \bm{K}(0)^2 \rangle = \frac{1}{h^2} \sum_{k=1}^N \sum_{l=1}^N \langle \bm{J}_k \cdot \bm{J}_l \rangle= \frac{1}{h^2} \left\langle \sum_{k=1}^N \bm{J}_k^2 \right\rangle ,
\end{split}
\end{align}
within the molecular-chaos assumption.
Instead of summing over particles, we can integrate the respective distribution over the colloid surface.
Consider an infinitesimal surface element of the spherical colloid of area $dS=R^2 \sin \theta d \theta d \varphi$.
During a time step, solvent particles with a relative velocity
\begin{align}
    \bar{\bm v}={\bm v}-{\bm u}-{\bm \Omega} \times R {\bm n},
\end{align}
which is negative in the normal direction, i.e., $\bar{\bm{v}} \cdot \bm{n}= \bar{v}_n < 0$, collide with the colloid
if they are located in the volume element $dV=-\bar{v}_n h dS$.
With the average particle density $n=N/V$, we find for the number $dN$ of colliding particles per surface element $dS$
\begin{align}
 dN=-n \bar{v}_n h dS
\end{align}
and consequently for the force autocorrelation function
\begin{align}
\langle {\bm K}(0)^2 \rangle= \frac{1}{h^2} \int dS \left\langle \frac{dN}{dS} {\bm J}^2 \right\rangle .
\end{align}
The dependence of $\bm{J}$ on $\bar{\bm{v}}$ is specified in Eq.(\ref{MomentumTransferCollisionsColloid}), and
the average is defined as
\begin{align}
\langle \dots \rangle=\int_{-\infty}^0 d\bar{v}_n \int_{-\infty}^\infty d\bar{v}_\theta \int_{-\infty}^\infty d\bar{v}_\varphi P_n(\bar{v}_n) P_t(\bar{v}_\theta) P_t(\bar{v}_\varphi) \dots,
\end{align}
where $\bar{v}_\theta$ and $\bar{v}_\varphi$ are the components of the tangential velocity $\bar{\bm{v}}_t$ in $\theta$ and $\varphi$ direction.
The probability distribution functions $P_n$ and $P_t$ are Gaussian,
since $\bm{v}, \bm{u}$, and $\bm{\Omega}$ are Maxwell-Boltzmann distributed.
The variance of $\bar{v}_n$ is $\sigma_n^2=k_BT/m+k_BT/M$, while the variance of $\bar{v}_\theta$ as well as $\bar{v}_\varphi$ is $\sigma_t^2=k_B T R^2/I+k_BT/m+k_BT/M$, where $I= \chi M R^2$.
Evaluation of the average yields
\begin{align}
 \langle {\bm K}(0)^2 \rangle= \frac{16}{h} \sqrt{2 \pi (k_B T)^3 \mu} n R^2 \frac{1+(2-\Gamma)\chi M/\mu}{1+\chi M/\mu}.
\end{align}
The friction coefficient, which we will denote as Enskog friction, follows by integration of the force autocorrelation function according to Eq.~(\ref{Eq:IntegrationFACFequalsGAMMA}).
Since MPC is a discrete-time-random process, the integral is
\begin{align}
  \int_0^\infty dt \langle \bm{K}(t) \cdot \bm{K}(0) \rangle=\frac{h}{2} \langle {\bm K}(0)^2 \rangle+h \sum_{l=1}^\infty \langle {\bm K}(l h) \cdot {\bm K}(0) \rangle.
\end{align}
Within the molecular chaos approximation, $\langle {\bm K}(l h) \cdot {\bm K}(0) \rangle=0$ for $l \neq 0$, and we find the Enskog friction coefficient
\begin{align}
   \gamma_E=\frac{8}{3} \sqrt{2 \pi k_\textrm{B} T \mu} n R^2 \frac{1+(2-\Gamma)\chi M/\mu}{1+\chi M/\mu},
\end{align}
In the time continuum limit, the force autocorrelation function is a delta distribution
\begin{align}
   \langle \bm{K}(t) \cdot \bm{K}(0) \rangle =  6 k_\textrm{B} T \gamma_E \delta(t),
\end{align}
which leads to the exponentially decaying velocity-autocorrelation function
\begin{align}
  C_u(t)= \frac{k_B T}{M} e^{- \gamma_E t/M}.
\end{align}
The rotational motion is treated in a similar manner. For the torque autocorrelation function at time $t=0$ we find
\begin{align}
\langle {\bm N}(0)^2 \rangle&= \int d\Omega \left\langle \frac{dN}{d\Omega}(R \bm{n} \times \bm{J})^2 \right\rangle /h^2  \\
&= \frac{16 n R^4}{h} \sqrt{2 \pi (k_B T)^3 \mu} (1- \Gamma) \frac{M \chi}{\mu+M \chi},
\end{align}
and hence, for the rotational friction coefficient
\begin{align}
  \xi_E=\frac{8}{3}\sqrt{2 \pi k_B T \mu} n R^4 (1- \Gamma) \frac{M \chi}{\mu+M \chi}.
\end{align}


%

\end{document}